\documentclass[aps,prd,twocolumn,floatfix,nofootinbib,superscriptaddress,longbibliography]{revtex4-2}

\usepackage{amsmath,amssymb}
\usepackage{booktabs}
\usepackage{graphicx}
\usepackage{hyperref}
\hypersetup{hidelinks}
\usepackage{xurl}
\usepackage{placeins}
\usepackage{array}
\begin{document}
	
	\title{Artifact-Conditioned Interval Diagnostics for Flow-Matching Neural Posterior Estimation in a Controlled Gravitational-Wave Benchmark}
	\author{Zhi Luo}
	\email[Corresponding author: ]{zhiluo@cqu.edu.cn}
	\affiliation{Department of Physics, Chongqing University, Chongqing 401331, China}
	\author{Qi-Qin Jing}
	\affiliation{School of Physics and Astronomy, China West Normal University, Nanchong 637002, China}
	\date{\today}
	
	\begin{abstract}
		Calibration checks for neural posterior estimators in gravitational-wave
		inference should remain interpretable when observations contain data-quality
		artifacts. We study marginal interval calibration in a controlled
		frequency-domain binary-black-hole benchmark with synthetic glitches,
		frequency masks, and power spectral density (PSD) mismatch. The posterior
		sampler is a support-aware flow-matching posterior estimator (FMPE) with a
		circular representation of coalescence phase. We compare raw marginal credible
		intervals with global rescaling, oracle artifact-stratified rescaling, hard
		predicted-label rescaling, and soft learned artifact-aware interval rescaling
		(LAIR). In the 1024-bin evaluation, a single global scale fitted on mixed
		calibration data transfers poorly to frequency-mask cases, giving a mean
		absolute 90\% marginal coverage error (MA90CE) of 0.1195. Soft LAIR lowers
		the corresponding error to 0.0672, although it is not uniformly better than
		the raw FMPE intervals. A 40-split LAIR evaluation and a six-checkpoint FMPE
		training-seed study show that the frequency-mask behavior is not a
		single-split artifact. The classifier
		recognizes frequency masks and PSD mismatch reliably, while glitch recall
		remains low. Waveform-resolution tests, PyCBC/LAL TaylorF2 backend checks,
		prior and Gaussian baselines, and controlled-likelihood reference-posterior
		probes indicate that marginal coverage must be read together with posterior
		width, geometry, and likelihood-based diagnostics. These results support
		using LAIR as an artifact-structured interval diagnostic, not as a substitute
		for posterior validation.
	\end{abstract}
	
	\maketitle
	
	\section{Introduction}
	
	Gravitational-wave parameter estimation infers compact-binary source
	parameters from noisy detector strain. Likelihood-based analyses provide the
	standard reference for precision inference, but repeated waveform evaluations
	and stochastic sampling can be costly. Neural posterior estimation reduces this
	cost by amortizing the map from strain data to posterior samples.
	
	Good calibration on clean simulations is not sufficient when the data
	distribution changes. Detector data can include nonstationary noise, missing
	frequency bands, glitches, power spectral density (PSD) drift, calibration
	uncertainty, and other effects absent from an idealized training set. Such
	effects can produce
	artifact-specific coverage errors. At the same time, acceptable marginal
	coverage is not by itself evidence of an informative posterior, since broad
	intervals can cover the truth while carrying little useful structure.
	
	We use a controlled frequency-domain binary-black-hole benchmark to isolate
	these effects. The simulator has a TaylorF2-like nonspinning waveform,
	analytic whitening, two detector channels, and known synthetic artifact labels.
	The posterior estimator is a conditional flow-matching posterior estimator
	(FMPE). Bounded parameters are trained in stable logit coordinates, and
	coalescence phase is represented by sine and cosine coordinates to avoid the
	artificial $0$--$2\pi$ discontinuity.
	
	The diagnostic considered here is learned artifact-aware interval rescaling
	(LAIR). Calibration events are used to fit marginal interval scale factors for
	each artifact family and parameter. At evaluation time an artifact classifier
	returns probabilities $p_\psi(a\mid x)$, and the deployed scale for parameter
	$k$ is the probability-weighted average of the artifact-specific scales. The
	procedure uses predicted artifact information for evaluation events; it does
	not require simulator labels for those events.
	
	The analysis asks whether artifact-conditioned interval scaling exposes
	failure modes that are hidden by a single global calibration. We evaluate the
	main 1024-bin FMPE checkpoint, repeat the LAIR calculation across 40
	independent simulator/calibration splits, and assess six independently trained
	1024-bin FMPE checkpoints. Classifier comparisons, waveform-resolution tests,
	simple posterior baselines, and controlled-likelihood reference probes are
	included so that the coverage results are not interpreted in isolation.
	
	\begin{figure*}[t]
		\centering
		\includegraphics[width=\textwidth]{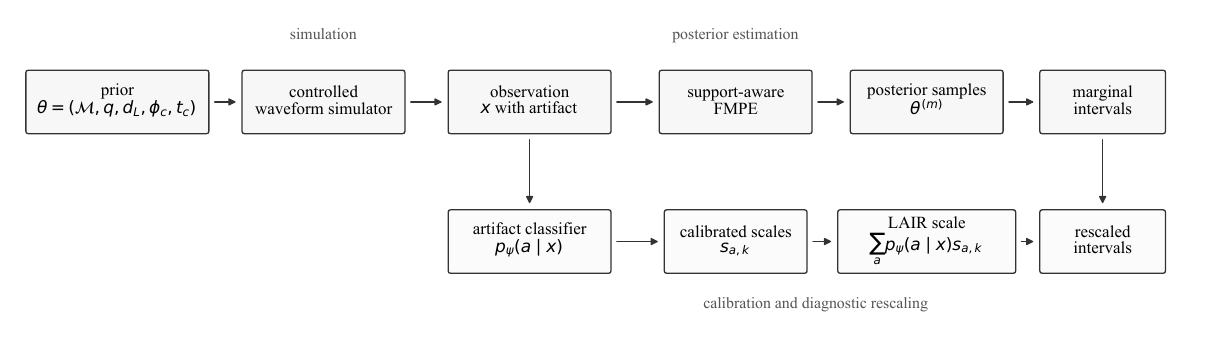}
		\caption{Inference and diagnostic workflow. Simulated observations are passed
			to a support-aware flow-matching posterior estimator. Marginal intervals are
			then evaluated raw or after global, oracle artifact-stratified, hard
			predicted-label, or soft LAIR rescaling.}
		\label{fig:pipeline-lair}
	\end{figure*}
	
	Figure~\ref{fig:pipeline-lair} gives the workflow. FMPE produces posterior
	samples, after which the LAIR-family procedures rescale marginal intervals and
	report artifact-conditioned coverage.
	
	\subsection{Relation to Prior Work}
	
	Neural posterior estimation has become an important route to faster
	gravitational-wave inference. Autoregressive normalizing flows showed that
	neural density estimators can learn compact-binary posteriors in controlled
	settings \cite{Green2020Flows}. DINGO-style systems extended this approach to
	rapid gravitational-wave analyses and added importance-sampling corrections
	for amortized proposals \cite{Dax2021Dingo,Dax2023DingoIS}. Flow matching
	formulates generation through a learned continuous vector field
	\cite{Lipman2022FM}, and has been adapted to simulation-based inference
	\cite{Dax2023FMPE}. The estimator used here belongs to this flow-matching
	family, with the artifact label available only through the controlled
	simulator.
	
	Posterior validation is a separate issue. Simulation-based calibration tests
	rank uniformity under the assumed simulator \cite{Talts2018SBC}, and
	TARP-style diagnostics provide related sampling-based checks
	\cite{Lemos2023TARP}. Marginal credible-interval coverage is therefore used
	here as a focused diagnostic. It can reveal artifact-dependent
	miscalibration, but it does not determine the joint posterior, tails, or
	information content.
	
	Likelihood-based parameter-estimation tools such as Bilby and PyCBC Inference
	remain the standard reference implementations for production analyses
	\cite{Ashton2019Bilby,Biwer2019PyCBC}; LALSuite supplies core waveform and
	data-analysis infrastructure \cite{Wette2020SWIGLAL}. Recent
	simulation-based studies have examined robustness to real noise and missing
	information \cite{Raymond2024RealNoise,Mao2025Gaps}, and public
	LIGO-Virgo-KAGRA (LVK) open data make such questions directly testable
	\cite{LVK2023O3OpenData,LVK2026O4bOpenData}. The experiments below do not use
	LVK strain data. They instead use known artifact families so that
	mode-resolved calibration errors can be measured directly.
	
	\subsubsection{Position relative to DINGO, FMPE, and likelihood-corrected neural inference}
	
	DINGO is the closest established neural posterior-estimation benchmark family
	for gravitational-wave inference. Its importance-sampling correction also
	illustrates how an amortized proposal can be checked against a likelihood. The
	present work does not implement that correction and does not run DINGO, Bilby,
	or PyCBC inference. The comparison is therefore limited in scope: LAIR is
	tested as a marginal artifact-conditioned interval diagnostic within a shared
	controlled simulator. The conditional RealNVP result in
	Sec.~\ref{sec:conditional-flow-baseline} is included only as a lightweight
	discrete-flow comparator under the same simulator and metrics.
	
	\subsubsection{Model-family choice: discrete flows, flow matching, and score-based diffusion}
	
	Discrete normalizing flows are the established neural-density baseline for
	gravitational-wave parameter estimation. DINGO-style conditional flows show
	that amortized density estimators can produce rapid compact-binary posterior
	samples. The RealNVP baseline below tests this discrete-flow family only within
	our simulator; it is not a reproduction of DINGO or DINGO-IS.
	
	The main estimator uses flow matching. Rather than applying a finite sequence
	of invertible maps, the model learns a continuous vector field that transports
	a simple base distribution to the observation-conditioned posterior target.
	This formulation is compatible with simulation-based inference and with the
	support-aware and circular parameterization used in the benchmark.
	
	Score-based diffusion is a natural alternative for conditional posterior
	sampling, but it is outside the present comparison. A fair diffusion study
	would require a separate sampler, runtime tuning, and calibration analysis
	under the same metrics.
	
	\section{Controlled Inference Problem and Methods}
	\label{sec:problem}
	
	\subsection{Parameters and Priors}
	
	The physical parameter vector is
	\[
	\theta=(\mathcal{M},q,d_L,\phi_c,t_c),
	\]
	where $\mathcal{M}$ is detector-frame chirp mass, $q=m_2/m_1\leq1$ is the
	mass ratio, $d_L$ is luminosity distance, $\phi_c$ is coalescence phase, and
	$t_c$ is coalescence time. The symmetric mass ratio and total mass are
	\[
	\eta=\frac{q}{(1+q)^2}, \qquad M=\mathcal{M}\eta^{-3/5}.
	\]
	In this paper $\mathcal{M}$ denotes chirp mass and $M$ denotes total mass. The
	priors are independent and uniform:
	\[
	\mathcal{M}\in[8,18]M_\odot,\quad
	q\in[0.55,1],\quad
	d_L\in[300,1800]\ {\rm Mpc},
	\]
	\[
	\phi_c\in[0,2\pi),\quad
	t_c\in[-0.035,0.035]\ {\rm s}.
	\]
	The prior volume is deliberately compact: it spans mass, distance, phase, and
	time uncertainty while keeping repeated single-GPU experiments feasible.
	
	\subsection{Signal and Observation Model}
	
	The simulator uses a transparent TaylorF2-like frequency-domain inspiral model
	rather than a production waveform approximant. The complex strain scales
	schematically as
	\[
	\tilde h(f;\theta)\propto
	\frac{\mathcal{M}^{5/6}}{d_L} f^{-7/6}\exp\{i\Psi(f;\theta)\},
	\]
	with phase contributions from coalescence time, coalescence phase, and a
	nonspinning post-Newtonian expansion. A smooth taper is applied near the
	inspiral cutoff. Each observation contains two detector channels represented
	by real and imaginary frequency-domain components after analytic whitening,
	together with a mask channel.
	
	This setup is a controlled inverse problem, not a full waveform model. It
	omits spins, precession, higher modes, sky localization, inclination,
	calibration uncertainty, and real detector nonstationarity. The restriction
	keeps the prior-generating process and artifact label known exactly. The
	PyCBC/LAL TaylorF2 comparison below checks phase, amplitude, and the fixed
	detector-projection conventions used in this simulator.
	
	\subsection{Controlled Likelihood and Whitening}
	
	The reference-posterior probes use the Gaussian likelihood implied by the
	whitened simulator. For detector channel $I$ and frequency bin $j$,
	\[
	d_{Ij}=m_j\left[\tilde h_I(f_j;\theta)+n_{Ij}\right],
	\]
	where $m_j\in\{0,1\}$ is the frequency mask and the whitened noise satisfies
	$n_{Ij}\sim\mathcal{N}(0,1)$ independently for real and imaginary components.
	The controlled log likelihood is
	\[
	\log p(d\mid\theta)
	=-\frac{1}{2}\sum_{I,j}m_j\,
	\left|d_{Ij}-\tilde h_I(f_j;\theta)\right|^2+{\rm const.}
	\]
	PSD factors are absorbed into the whitening operation. The two detector
	channels use fixed gains and time delays rather than sky-dependent antenna
	patterns, so the problem remains a controlled two-channel inverse problem
	rather than a production detector response model.
	
	PSD mismatch is introduced by changing the whitening tilt used to generate an
	evaluation observation while retaining the nominal whitening convention for
	the trained estimator and controlled reference model. Frequency-mask artifacts
	set a contiguous band of $m_j$ to zero and expose the missingness through the
	mask channel. Glitch artifacts add a localized sine-Gaussian-like disturbance
	after the clean signal is generated. These definitions make the artifact label
	and the likelihood mismatch known by construction.
	
	\subsection{Artifact Families}
	
	The label set is
	\[
	a\in\{\mathrm{clean},\mathrm{glitch},\mathrm{frequency\ mask},
	\mathrm{PSD\ mismatch}\}.
	\]
	Clean observations use the nominal simulator. Glitch observations contain a
	localized sine-Gaussian-like disturbance. Frequency-mask observations remove a
	band of frequency bins and expose the missingness through the mask channel.
	PSD-mismatch observations perturb the whitening model.
	
\begin{figure}[!htbp]
		\centering
		\includegraphics[width=0.72\linewidth]{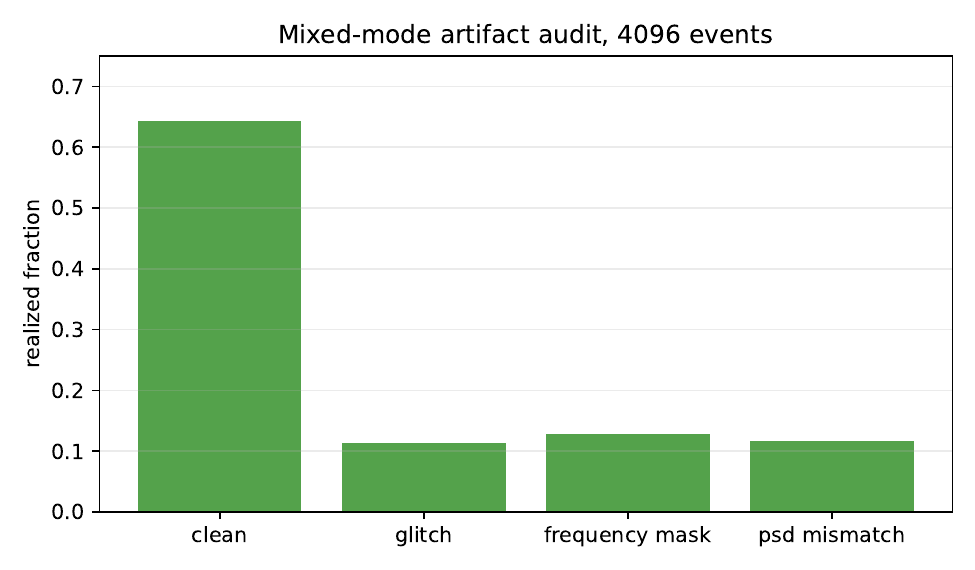}
		\caption{Realized artifact counts in a 4096-event mixed-mode sample. The clean
			fraction follows the configured 0.65 probability, and the artifact families are
			approximately balanced within the remaining probability mass.}
		\label{fig:artifact-audit}
	\end{figure}

\begin{figure*}[t]
		\centering
		\includegraphics[width=0.82\textwidth]{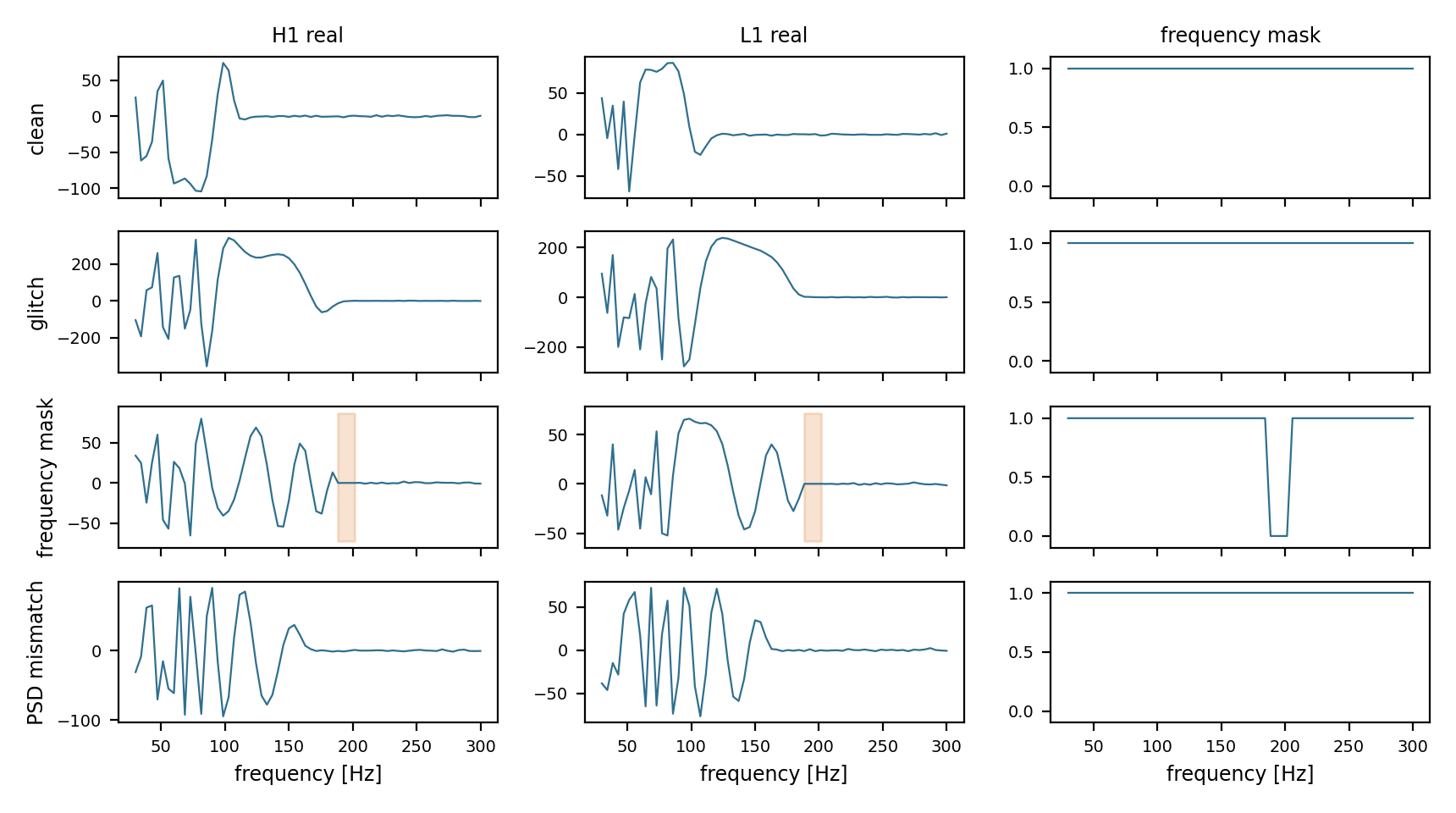}
		\caption{Representative synthetic artifact examples. Frequency masks affect
			the mask channel directly; glitches and PSD mismatch appear as subtler
			frequency-dependent structure.}
		\label{fig:artifact-examples}
	\end{figure*}

	In mixed evaluation, a clean observation is drawn with probability 0.65 and an
	artifact observation with probability 0.35. Conditional on an artifact, the
	three artifact families are sampled uniformly. A 4096-event sample gives the
	realized fractions in Table~\ref{tab:artifact-audit} and
	Fig.~\ref{fig:artifact-audit}. Qualitative examples appear in
	Fig.~\ref{fig:artifact-examples}.
	
	\begin{table}[!htbp]
\centering
\caption{Realized mixed-mode artifact composition from 4096 simulated events.}
\label{tab:artifact-audit}
\begin{tabular}{lrr}
\toprule
Artifact & Count & Fraction \\
\midrule
clean & 2632 & 0.6426 \\
glitch & 463 & 0.1130 \\
frequency mask & 522 & 0.1274 \\
PSD mismatch & 479 & 0.1169 \\
\bottomrule
\end{tabular}
\end{table}

	\subsection{Posterior Estimator and Calibration Diagnostics}
	\label{sec:methods}
	
	\subsubsection{Support-Aware Transforms}
	
	For each bounded scalar parameter $x\in[\ell,u]$, the model uses
	\[
	z=\frac{x-\ell}{u-\ell}, \qquad
	y=\log\frac{z}{1-z},
	\]
	with numerical clipping before the logit. The inverse transform maps samples
	back into the physical prior interval. This keeps decoded $\mathcal{M}$, $q$,
	$d_L$, and $t_c$ samples inside the prior support. Unless stated otherwise,
	interval rescaling for bounded scalar parameters is performed in this
	transformed coordinate: endpoints are mapped through the box logit, widened or
	narrowed about the transformed posterior center, and decoded back to the
	physical interval. This convention avoids unnecessary asymmetry near hard
	prior boundaries.
	
	\subsubsection{Circular Phase Treatment}
	
	Coalescence phase is embedded as
	\[
	(\cos\phi_c,\sin\phi_c)
	\]
	during training and decoded with $\mathrm{atan2}$ modulo $2\pi$. The model
	target is therefore
	\[
	y(\theta)=
	\left(
	\mathrm{logit}\,\mathcal{M},
	\mathrm{logit}\,q,
	\mathrm{logit}\,d_L,
	\cos\phi_c,
	\sin\phi_c,
	\mathrm{logit}\,t_c
	\right).
	\]
	The phase coordinates are decoded after normalization by
	$r_\phi=\sqrt{u_\phi^2+v_\phi^2}$, where $(u_\phi,v_\phi)$ is the generated
	phase embedding. The implementation records $r_\phi$ as an off-circle
	diagnostic. Large departures from unity indicate that the Euclidean flow has
	moved away from the training manifold even when the decoded angle is defined.
	Evaluation uses circular phase errors and geodesic intervals on $S^1$. The
	wrapped phase residual is
	\[
	\Delta\phi=\mathrm{atan2}\left(\sin(\phi-\phi^\star),
	\cos(\phi-\phi^\star)\right).
	\]
	For an interval with circular center $\bar\phi$ and half-width $h\leq\pi$, the
	truth is covered when
	\[
	\left|\mathrm{atan2}\left(\sin(\phi^\star-\bar\phi),
	\cos(\phi^\star-\bar\phi)\right)\right|\leq h .
	\]
	This definition removes the discontinuity at the $0$--$2\pi$ boundary.
	
	\subsubsection{Conditional Flow Matching}
	
	Let $x$ denote the observation and $y$ the support-aware target. A base sample
	$z_0$ is drawn from a standard Gaussian. Conditional flow matching trains a
	time-dependent vector field $v_\varphi(z,t,x)$ using the straight-line
	interpolation
	\[
	z_t=(1-t)z_0+t y,
	\]
	with target velocity $y-z_0$. The objective is
	\[
	\mathcal{L}(\varphi)=
	\mathbb{E}_{x,y,z_0,t}
	\left[
	\left\|v_\varphi(z_t,t,x)-(y-z_0)\right\|_2^2
	\right].
	\]
	At inference time the learned ordinary differential equation is integrated
	from the Gaussian base distribution to $t=1$, and generated samples are decoded
	to normalized physical coordinates for evaluation.
	
	\subsubsection{Marginal Interval Rescaling}
	
	Raw posterior samples define the marginal credible intervals. Global rescaling
	fits one scale factor per parameter on calibration data and applies it to all
	evaluation intervals. Oracle artifact-stratified rescaling fits one scale per
	artifact and parameter, then uses the true artifact label. Predicted-label
	rescaling replaces the true label with the classifier's hard prediction. These
	operations affect marginal interval widths only; they do not move posterior
	centers or reconstruct joint correlations. For artifact class $a$, parameter
	$k$, and nominal levels $\alpha\in\mathcal{A}$, the calibrated scale is
	\[
	s_{a,k}=
	\arg\min_{s>0}
	\sum_{\alpha\in\mathcal{A}} w_\alpha
	\left|
	\hat c_{\rm cal}(a,k,\alpha;s)-\alpha
	\right|,
	\]
	where $\hat c_{\rm cal}$ is empirical calibration coverage after multiplying
	the marginal half-width by $s$. Global rescaling uses the same objective after
	pooling artifact classes. The primary scalar metric below is based on
	$\alpha=0.9$, while coverage curves evaluate the fitted intervals over a grid
	of nominal levels.
	
	\subsubsection{Artifact Classifier and LAIR}
	
	The artifact classifier is a one-dimensional convolutional network over the
	frequency axis. Its channels are detector real and imaginary components plus
	the mask. Let $p_\psi(a\mid x)$ be the classifier probability for artifact
	class $a$. LAIR fits scale factors $s_{a,k}$ for artifact class $a$ and
	parameter $k$ on calibration data. The soft deployed scale is
	\[
	s_k(x)=\sum_a p_\psi(a\mid x)s_{a,k}.
	\]
	The rescaled marginal interval is centered on the event's posterior center,
	with its half-width multiplied by $s_k(x)$.
	
	This implementation of LAIR is an artifact-conditioned marginal interval
	diagnostic. It rescales intervals but leaves posterior centers and joint sample
	geometry unchanged, so its output must be interpreted together with the raw
	samples and reference diagnostics.
	
	\subsubsection{Baselines}
	
	Two baselines contextualize the coverage results. A prior-only baseline asks
	how often the truth falls in central prior intervals. A diagonal Gaussian
	posterior baseline gives a simple parametric approximation in the 256-bin
	low-resolution setting. Low MA90CE for either baseline does not imply an
	accurate posterior; it can also reflect broad intervals.
	
	\subsubsection{Metrics}
	
	For nominal level $\alpha$, the empirical marginal coverage for parameter $k$
	is
	\[
	\hat c_k(\alpha)=\frac{1}{N}\sum_{i=1}^N
	\mathbf{1}\{\theta_{i,k}^{\rm true}\in I_{i,k}^{\alpha}\}.
	\]
	The primary scalar metric is mean absolute 90\% marginal coverage error,
	\[
	\mathrm{MA90CE}=\frac{1}{K}\sum_{k=1}^K|\hat c_k(0.9)-0.9|.
	\]
	We also track normalized mean absolute error, bias, posterior widths, and
	coverage curves. Event-bootstrap intervals resample evaluation events for a
	fixed trained model and fixed calibration procedure.
	
	\section{Experiments and Results}
	\label{sec:setup}
	\label{sec:results}
	
	All experiments use the controlled simulator described above. The FMPE tiers are
	summarized in Table~\ref{tab:experiment-config}. The 1024-bin tier is the main
	high-resolution checkpoint. The 2048-bin tier tests frequency resolution at a
	smaller evaluation budget. The 256-bin tier is retained as a low-resolution
	check for earlier LAIR and baseline diagnostics.
	
	\begin{table*}[t]
\centering
\caption{Support-aware FMPE experiment tiers. Evaluation counts are per artifact mode; provenance details are retained in the reproducibility records.}
\label{tab:experiment-config}
\small
\begin{tabular}{lrrrrr}
\toprule
Tier & Bins & Events & Samples & ODE steps & Train steps \\
\midrule
low-resolution check & 256 & 12 & 96 & 16 & 80 \\
intermediate check & 512 & 64 & 256 & 32 & 600 \\
primary 1024-bin study & 1024 & 256 & 512 & 48 & 2000 \\
2048-bin resolution study & 2048 & 128 & 256 & 48 & 1500 \\
\bottomrule
\end{tabular}
\end{table*}

	The 1024-bin LAIR evaluation uses the trained 1024-bin checkpoint, a matching
	1024-bin artifact classifier, 256 mixed calibration events, 256 evaluation
	events per mode, 512 posterior samples per event, 48 ODE steps, and 200
	bootstrap resamples over evaluation events. A repeated evaluation reuses the
	same FMPE checkpoint across 40 independent simulator/calibration splits. That
	run measures evaluation-split and calibration-split variability; it is not a
	multi-seed FMPE training study. The classifier is trained on balanced
	simulated artifact batches and evaluated on 2048 balanced validation events.
	All figures and tables are generated from retained metric files.
	
	\subsection{Waveform-Resolution Diagnostic}
	
	The waveform-resolution diagnostic compares frequency grids from 64 to 2048
	bins with a 4096-bin reference over 30--300 Hz using 128 prior draws. The
	mismatch is maximized over constant phase but not over time shifts.
	
	\begin{table}[!htbp]
\centering
\caption{Waveform-resolution diagnostic against a 4096-bin reference over 30--300 Hz using 128 prior draws. Mismatch is phase-maximized but not time-shift-maximized.}
\label{tab:waveform-highres}
\begin{tabular}{rrrr}
\toprule
Bins & Relative L2 & Phase RMS [rad] & Mismatch \\
\midrule
64 & 0.562471 & 136.647816 & 0.177342 \\
128 & 0.395162 & 123.246203 & 0.084973 \\
256 & 0.236611 & 96.954127 & 0.031709 \\
512 & 0.105759 & 48.762003 & 0.007561 \\
1024 & 0.033538 & 3.370806 & 0.000847 \\
2048 & 0.008697 & 0.000163 & 0.000059 \\
\bottomrule
\end{tabular}
\end{table}

	\begin{figure}[!htbp]
		\centering
		\includegraphics[width=\linewidth]{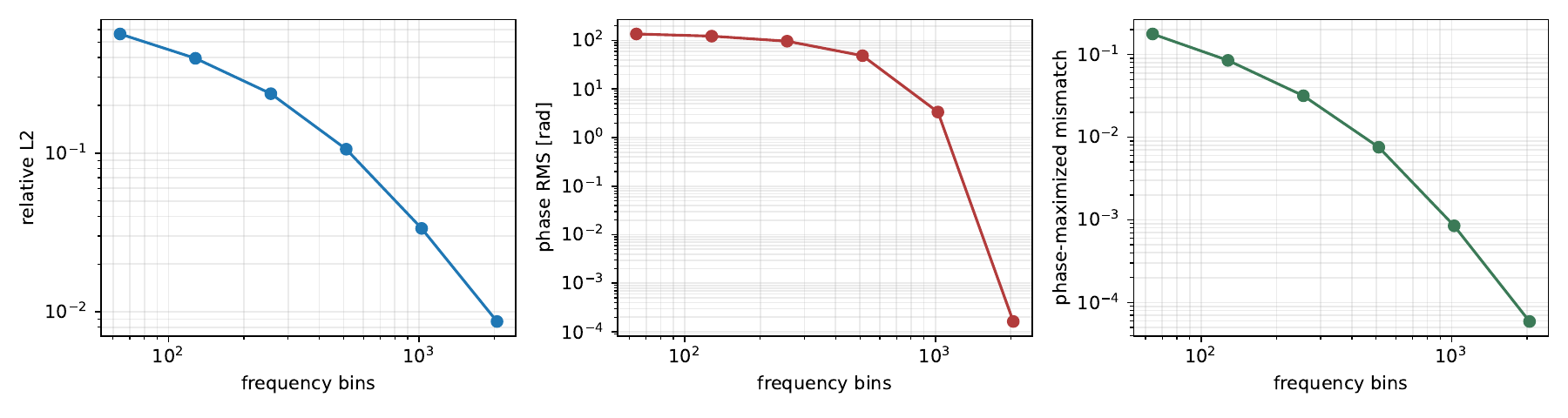}
		\caption{Waveform-resolution diagnostic against a 4096-bin reference. The
			64-, 128-, and 256-bin grids are visibly coarse, while the 1024- and 2048-bin
			grids are close to the reference under the phase-maximized mismatch used here.}
		\label{fig:waveform-highres}
	\end{figure}
	
	Table~\ref{tab:waveform-highres} and Fig.~\ref{fig:waveform-highres} show the
	effect of frequency resolution. The 256-bin grid has mismatch 0.0317 against
	the 4096-bin reference. The 1024-bin grid reduces the mismatch to
	$8.47\times10^{-4}$, and the 2048-bin grid reaches $5.89\times10^{-5}$. We
	therefore use the 256-bin results only for low-resolution checks, while the 1024-bin
	results carry the main inference analysis.
	
	\begingroup
\begin{center}
\refstepcounter{table}\label{tab:lalsim-validation}
{\scriptsize \textbf{TABLE~\thetable.} TaylorF2 backend checks against PyCBC/LAL routines. The controlled-response row uses the same fixed detector gains, delays, and analytic whitening as the simulator.\par}
\vspace{2pt}
\scriptsize
\setlength{\tabcolsep}{3pt}
\begin{tabular}{>{\raggedright\arraybackslash}p{0.32\columnwidth}>{\raggedright\arraybackslash}p{0.38\columnwidth}>{\raggedright\arraybackslash}p{0.20\columnwidth}}
\toprule
Check & Metric & Interpretation \\
\midrule
Intrinsic phase/amplitude & median phase RMS 0.0272 rad; median mismatch 0.00468 & consistent with controlled phase shape \\
Convention-aligned controlled response & median mismatch $5.23\times10^{-4}$; max mismatch 0.0185 & consistent with controlled response \\
Direct PyCBC sky projection & mean relative $L_2$ error 1.411; mean phase RMS 43.9 rad & out of scope \\
\bottomrule
\end{tabular}
\end{center}
\endgroup

	Table~\ref{tab:lalsim-validation} summarizes the TaylorF2 backend checks. In
	the intrinsic comparison, the PyCBC waveform is allowed to differ by arbitrary
	amplitude scale, constant phase, and time translation. After those
	convention-dependent factors are removed, the median phase residual is
	0.0272 rad and the median mismatch is 0.00468 over 20 prior draws. The
	controlled detector response then applies the same fixed gains, fixed delays,
	and analytic whitening as the simulator. In that like-for-like comparison, the
	median detector-response mismatch is $5.23\times10^{-4}$ and the maximum
	mismatch is 0.0185. A direct comparison with a PyCBC sky-response projection
	uses a different detector-response model and is not used as validation
	evidence. The simulator is therefore treated as a controlled TaylorF2-like
	inverse problem throughout the analysis.
	
	\subsection{High-Resolution FMPE Results}
	
	\begin{table}[!htbp]
\centering
\caption{Support-aware FMPE runs. MA90CE is mean absolute 90\% marginal coverage error with circular phase treatment. Global denotes marginal interval rescaling fitted on a mixed calibration set.}
\label{tab:arxiv-main}
\begin{tabular}{lrrrr}
\toprule
Mode & \multicolumn{2}{c}{1024 bins} & \multicolumn{2}{c}{2048 bins} \\
\cmidrule(lr){2-3}\cmidrule(lr){4-5}
 & Raw & Global & Raw & Global \\
\midrule
clean & 0.0305 & 0.0620 & 0.0372 & 0.0644 \\
glitch & 0.0458 & 0.0605 & 0.0428 & 0.0625 \\
frequency mask & 0.0369 & 0.1148 & 0.0559 & 0.1281 \\
PSD mismatch & 0.0436 & 0.0575 & 0.0453 & 0.0581 \\
mixed & 0.0319 & 0.0773 & 0.0272 & 0.0734 \\
\bottomrule
\end{tabular}
\end{table}

	\begin{figure}[!htbp]
		\centering
		\includegraphics[width=0.82\linewidth]{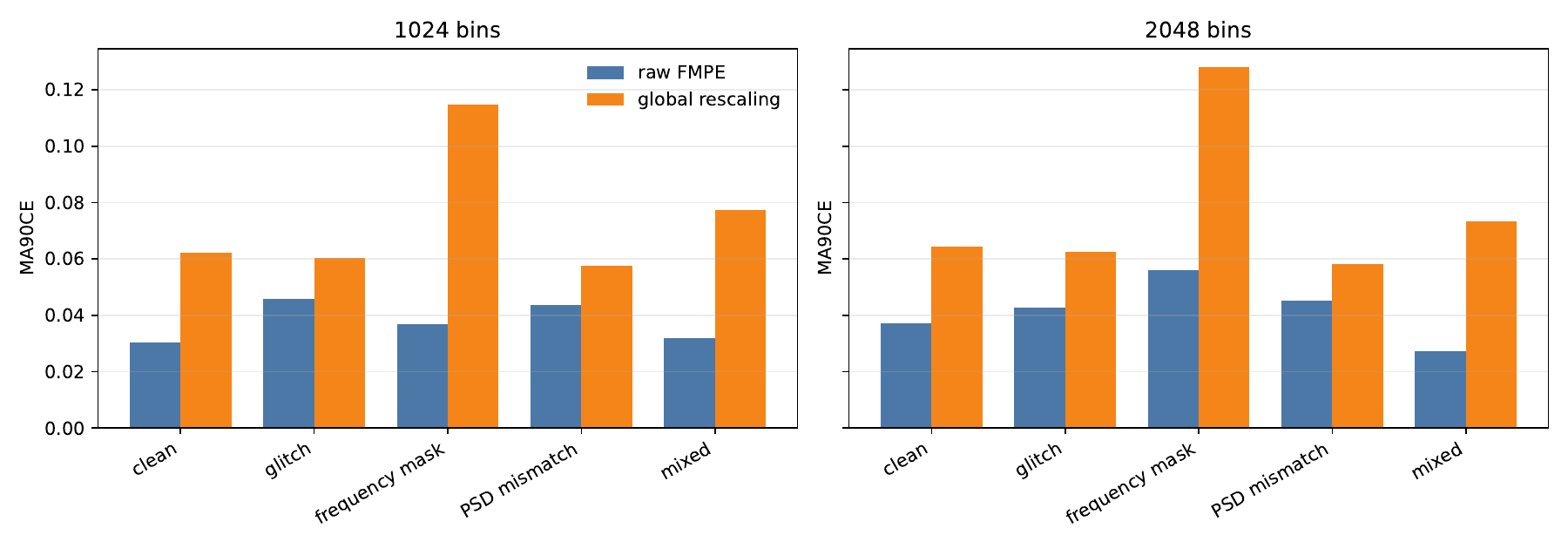}
		\caption{Raw and globally rescaled MA90CE for the trained 1024-bin and 2048-bin
			FMPE evaluations. Global marginal rescaling is not reliably beneficial and is
			especially poor under frequency masking.}
		\label{fig:highres-method-comparison}
	\end{figure}
	
	The high-resolution FMPE evaluations are summarized in
	Table~\ref{tab:arxiv-main} and Fig.~\ref{fig:highres-method-comparison}. At
	1024 bins, raw MA90CE ranges from 0.0305 in clean mode to 0.0458 in glitch
	mode. The 2048-bin extended run is similar, with raw MA90CE from 0.0272 to
	0.0559 across modes. In both tiers, global marginal rescaling increases
	MA90CE for every reported mode. The frequency-mask mode is most affected:
	global rescaling gives MA90CE 0.1148 at 1024 bins and 0.1281 at 2048 bins.
	
	A scale fitted on mixed calibration data therefore does not transfer uniformly
	across artifact modes. It can reduce an aggregate coverage error while
	worsening a particular data-quality condition; in these runs, frequency
	masking is the most visible case.
	
	\begin{table}[!htbp]
\centering
\caption{Standardized 1024-bin FMPE training-seed study across six checkpoints. Values are mean MA90CE with seed-to-seed standard deviation in parentheses. Each checkpoint is evaluated with 256 calibration events, 256 evaluation events per mode, 512 posterior samples per event, and 48 ODE steps. Global denotes marginal interval rescaling fitted on the calibration set.}
\label{tab:fmpe-seed-sensitivity}
\scriptsize
\begin{tabular}{lrr}
\toprule
Mode & Raw FMPE & Global \\
\midrule
clean & 0.0421 (0.0093) & 0.0402 (0.0158) \\
glitch & 0.0448 (0.0092) & 0.0362 (0.0167) \\
frequency mask & 0.0443 (0.0128) & 0.1175 (0.0271) \\
PSD mismatch & 0.0406 (0.0038) & 0.0397 (0.0153) \\
mixed & 0.0334 (0.0082) & 0.0393 (0.0222) \\
\bottomrule
\end{tabular}
\end{table}

	Table~\ref{tab:fmpe-seed-sensitivity} reports a standardized 1024-bin
	training-seed study across six checkpoints: the main checkpoint, two earlier
	additional seeds, and three additional checkpoints trained with the same
	2000-step protocol. All six are evaluated with 256 calibration events and 256
	evaluation events per artifact mode. The study supports the frequency-mask
	conclusion:
	raw FMPE has mean frequency-mask MA90CE 0.0443 with seed-to-seed standard
	deviation 0.0128, whereas global rescaling gives 0.1175 with standard
	deviation 0.0271. The mixed-mode global result also varies more than the raw
	result. This study probes training-initialization sensitivity within the
	controlled benchmark; it is not a substitute for larger production-scale
	retraining studies.
	
	\subsection{Conditional Normalizing-Flow Baseline}
	\label{sec:conditional-flow-baseline}
	
	\begin{table}[!htbp]
\centering
\caption{1024-bin conditional RealNVP baseline compared with the trained FMPE checkpoint. This is a controlled discrete-flow baseline in the same simulator, not a DINGO reproduction. Entries are MA90CE.}
\label{tab:conditional-flow-baseline}
\small
\begin{tabular}{lrrrr}
\toprule
Mode & \multicolumn{2}{c}{FMPE} & \multicolumn{2}{c}{Conditional RealNVP} \\
\cmidrule(lr){2-3}\cmidrule(lr){4-5}
 & Raw & Global & Raw & Global \\
\midrule
clean & 0.0305 & 0.0620 & 0.0362 & 0.0541 \\
glitch & 0.0458 & 0.0605 & 0.0425 & 0.0806 \\
frequency mask & 0.0369 & 0.1148 & 0.0728 & 0.1141 \\
PSD mismatch & 0.0436 & 0.0575 & 0.0303 & 0.0419 \\
mixed & 0.0319 & 0.0773 & 0.0347 & 0.0619 \\
\bottomrule
\end{tabular}
\end{table}

	Table~\ref{tab:conditional-flow-baseline} adds a 1024-bin conditional RealNVP
	baseline trained and evaluated in the same simulator. The baseline uses an
	invertible box-logit phase coordinate for density modeling and decodes phase
	circularly for metrics. RealNVP is close to FMPE in clean and mixed raw
	MA90CE, worse under frequency masking, and better in this single run for PSD
	mismatch. Both model families are sensitive to global rescaling, so the
	artifact dependence is not unique to the FMPE sampler.
	
	\subsection{Representative Posterior Geometry}
	
	\begin{figure*}[t]
		\centering
		\includegraphics[width=0.72\textwidth]{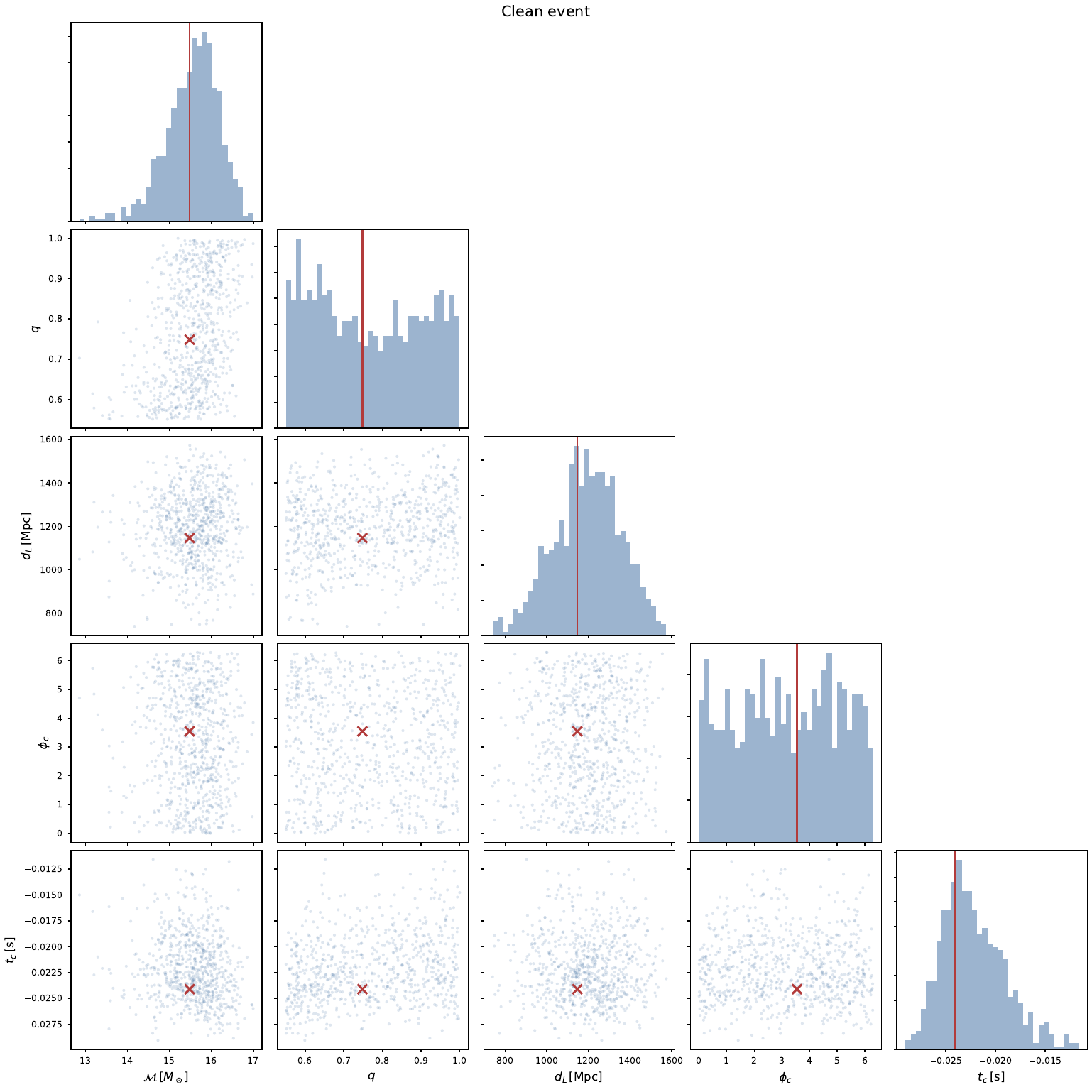}
		\caption{Representative 1024-bin FMPE posterior geometry for one clean
			simulated event. The corner plot uses the trained 1024-bin checkpoint and 768
			posterior samples. It shows marginal and pairwise structure; no reference
			posterior is overlaid.}
		\label{fig:corner-1024-clean}
	\end{figure*}
	
	\begin{figure*}[t]
		\centering
		\includegraphics[width=0.72\textwidth]{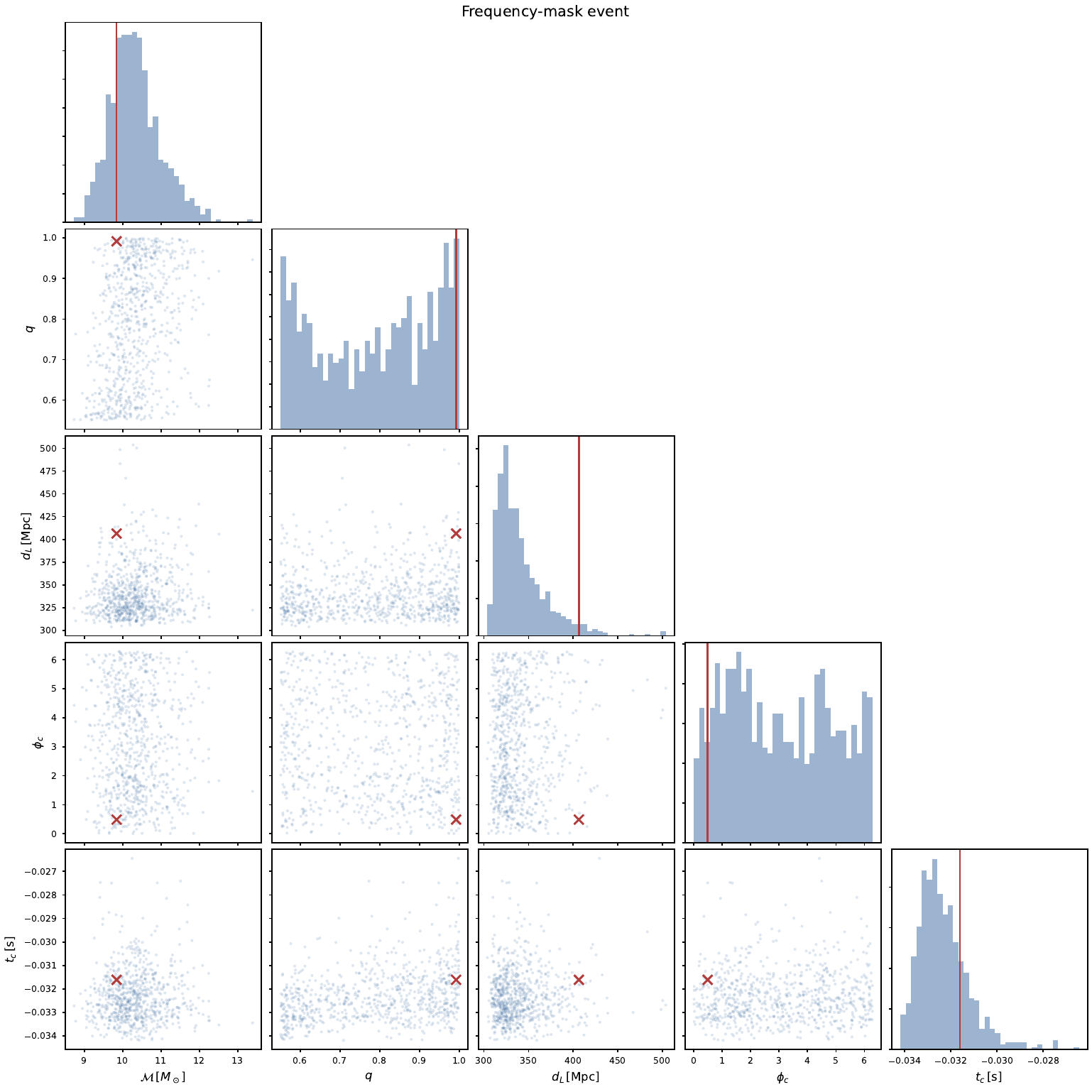}
		\caption{Representative 1024-bin FMPE posterior geometry for one frequency-mask
			simulated event, using the same plotting convention as
			Fig.~\ref{fig:corner-1024-clean}.}
		\label{fig:corner-1024-mask}
	\end{figure*}
	
	Figures~\ref{fig:corner-1024-clean} and \ref{fig:corner-1024-mask} show
	posterior geometry for the trained 1024-bin checkpoint. They are included
	because marginal coverage tables do not display correlation structure or
	multimodality. The likelihood-reference comparison is reported separately in
	Sec.~\ref{sec:reference-posterior}; in the checked events it shows substantial
	FMPE--reference differences.
	
	\subsection{1024-Bin LAIR Evaluation}
	
	\begin{table*}[t]
\centering
\caption{1024-bin LAIR evaluation using the trained 1024-bin FMPE checkpoint, a matching 1024-bin artifact classifier, 256 calibration events, 256 evaluation events per mode, 512 posterior samples per event, 48 ODE steps, and 200 bootstrap resamples. Values are MA90CE.}
\label{tab:lair-1024}
\begin{tabular}{lrrrrr}
\toprule
Mode & Raw FMPE & Global & Oracle & Predicted label & Soft LAIR \\
\midrule
clean & 0.0452 & 0.0331 & 0.0370 & 0.0447 & 0.0423 \\
glitch & 0.0484 & 0.0169 & 0.0705 & 0.0253 & 0.0273 \\
frequency mask & 0.0541 & 0.1195 & 0.0672 & 0.0672 & 0.0672 \\
PSD mismatch & 0.0397 & 0.0394 & 0.0634 & 0.0486 & 0.0634 \\
mixed & 0.0303 & 0.0414 & 0.0453 & 0.0422 & 0.0594 \\
\bottomrule
\end{tabular}
\end{table*}

	\begin{figure}[!htbp]
		\centering
		\includegraphics[width=0.86\linewidth]{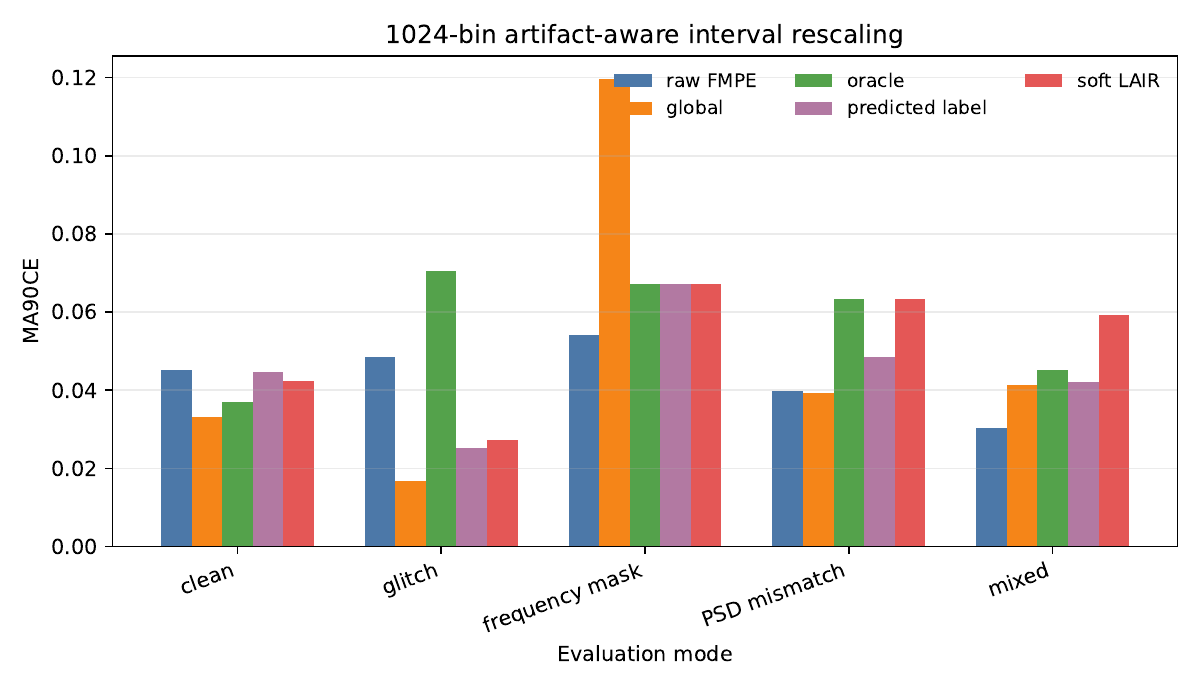}
		\caption{1024-bin LAIR comparison using the trained high-resolution FMPE
			checkpoint and a matching 1024-bin artifact classifier. Rescaling changes the
			coverage error in a strongly mode-dependent way.}
		\label{fig:lair-1024-comparison}
	\end{figure}
	
	Table~\ref{tab:lair-1024} and Fig.~\ref{fig:lair-1024-comparison} give the
	1024-bin artifact-conditioned result. Global rescaling improves clean MA90CE
	from 0.0452 to 0.0331 and glitch MA90CE from 0.0484 to 0.0169. Under
	frequency masking, however, the same global scale raises MA90CE from 0.0541 to
	0.1195. Soft LAIR gives 0.0273 in glitch mode and reduces the frequency-mask
	error to 0.0672, but it is worse than the raw FMPE intervals for PSD-mismatch
	and mixed mode.
	
	The oracle artifact-stratified result is also not uniformly beneficial. Even
	with true artifact labels, a marginal width correction can move coverage in
	the wrong direction when the fitted scale is noisy or when the dominant
	posterior error is not a width error. The method should therefore be read as a
	mode-resolved marginal coverage diagnostic, not as a correction to posterior
	sample geometry.
	
	\begin{table*}[t]
\centering
\caption{Event-bootstrap 95\% intervals for 1024-bin MA90CE. Each entry is estimate [2.5\%, 97.5\%].}
\label{tab:lair-1024-bootstrap}
\small
\begin{tabular}{llll}
\toprule
Mode & Raw FMPE & Global & Soft LAIR \\
\midrule
clean & 0.0452 [0.0334, 0.0567] & 0.0331 [0.0190, 0.0502] & 0.0423 [0.0267, 0.0610] \\
glitch & 0.0484 [0.0383, 0.0594] & 0.0169 [0.0136, 0.0369] & 0.0273 [0.0159, 0.0500] \\
frequency mask & 0.0541 [0.0356, 0.0806] & 0.1195 [0.0914, 0.1469] & 0.0672 [0.0470, 0.0930] \\
PSD mismatch & 0.0397 [0.0289, 0.0530] & 0.0394 [0.0268, 0.0560] & 0.0634 [0.0439, 0.0814] \\
mixed & 0.0303 [0.0186, 0.0475] & 0.0414 [0.0283, 0.0603] & 0.0594 [0.0416, 0.0767] \\
\bottomrule
\end{tabular}
\end{table*}

	The event-bootstrap intervals in Table~\ref{tab:lair-1024-bootstrap} quantify
	finite-evaluation uncertainty. For frequency masks, the raw, global, and
	soft-LAIR intervals are [0.0356, 0.0806], [0.0914, 0.1469], and [0.0470,
	0.0930], respectively. For mixed mode, soft LAIR gives MA90CE 0.0594 with
	interval [0.0416, 0.0767], whereas the raw estimate is 0.0303 with interval
	[0.0186, 0.0475].
	
	\begin{figure}[!htbp]
		\centering
		\includegraphics[width=\linewidth]{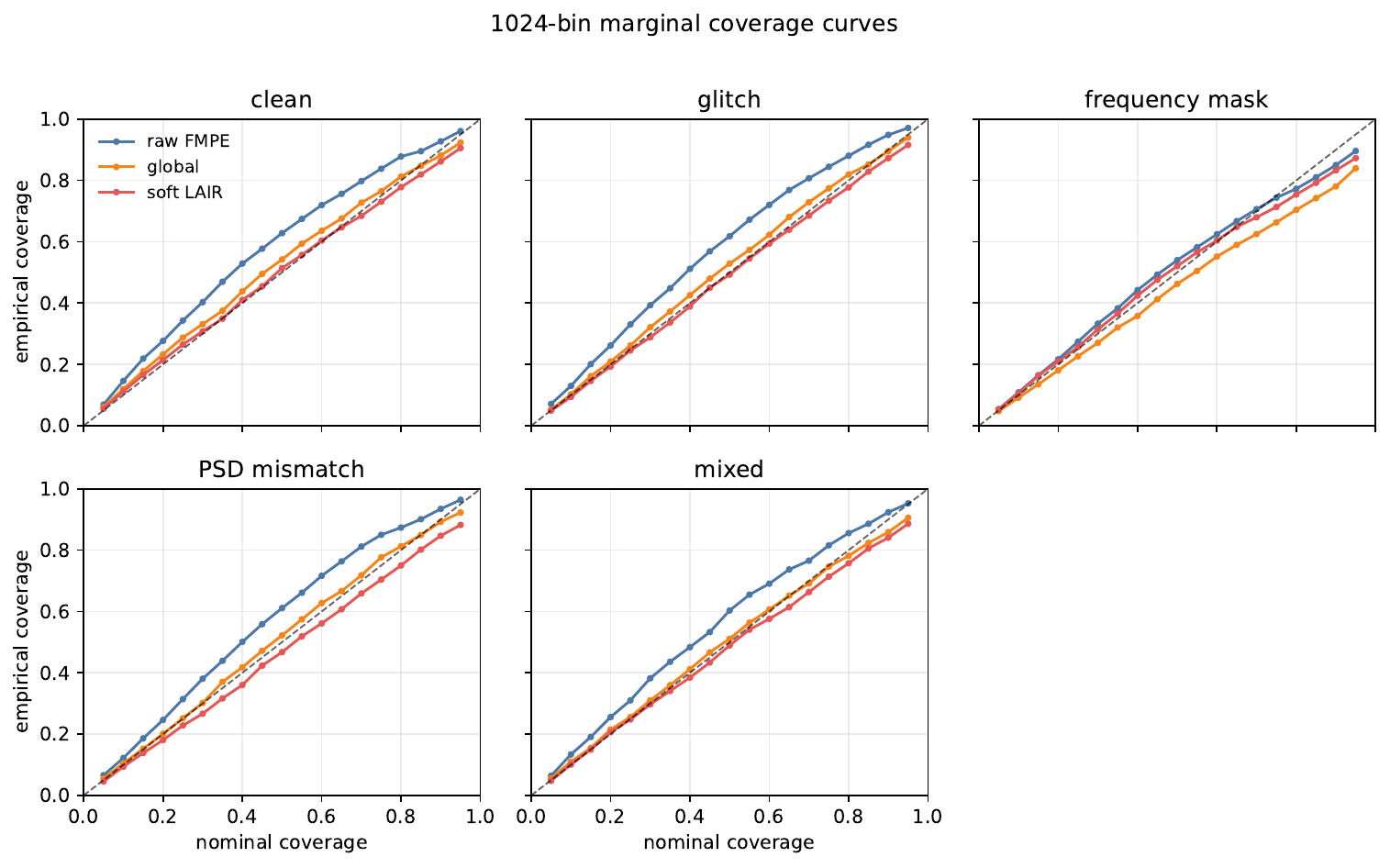}
		\caption{1024-bin marginal coverage curves averaged over parameters for raw
			FMPE, global rescaling, and soft LAIR. The diagonal is ideal calibration.
			Frequency masking shows the most visible separation between global rescaling
			and the artifact-aware diagnostic.}
		\label{fig:lair-1024-coverage}
	\end{figure}
	
	Figure~\ref{fig:lair-1024-coverage} shows the corresponding coverage curves.
	Rescaling affects nominal levels differently, and the ordering of methods
	changes across artifact modes.
	
	\begin{table*}[t]
\centering
\caption{Repeated 1024-bin LAIR calibration study across 40 independent simulator/calibration splits using the validation-selected classifier from the classifier comparison. Entries are mean MA90CE with split-to-split standard deviation in parentheses. This classifier is reported separately because it is not the higher-performing classifier in Table~\ref{tab:classifier-1024}.}
\label{tab:lair-1024-multiseed}
\scriptsize
\begin{tabular}{lrrrrr}
\toprule
Mode & Raw FMPE & Global & Oracle & Predicted label & Soft LAIR \\
\midrule
clean & 0.0433 (0.0059) & 0.0367 (0.0119) & 0.0394 (0.0121) & 0.0427 (0.0136) & 0.0425 (0.0142) \\
glitch & 0.0432 (0.0055) & 0.0353 (0.0085) & 0.0593 (0.0225) & 0.0401 (0.0108) & 0.0420 (0.0138) \\
frequency mask & 0.0521 (0.0096) & 0.1137 (0.0165) & 0.0679 (0.0202) & 0.0679 (0.0202) & 0.0679 (0.0201) \\
PSD mismatch & 0.0436 (0.0072) & 0.0344 (0.0089) & 0.0581 (0.0201) & 0.0386 (0.0107) & 0.0409 (0.0124) \\
mixed & 0.0323 (0.0068) & 0.0327 (0.0100) & 0.0380 (0.0111) & 0.0374 (0.0107) & 0.0403 (0.0123) \\
\bottomrule
\end{tabular}
\end{table*}

	\begin{figure}[!htbp]
		\centering
		\includegraphics[width=0.86\linewidth]{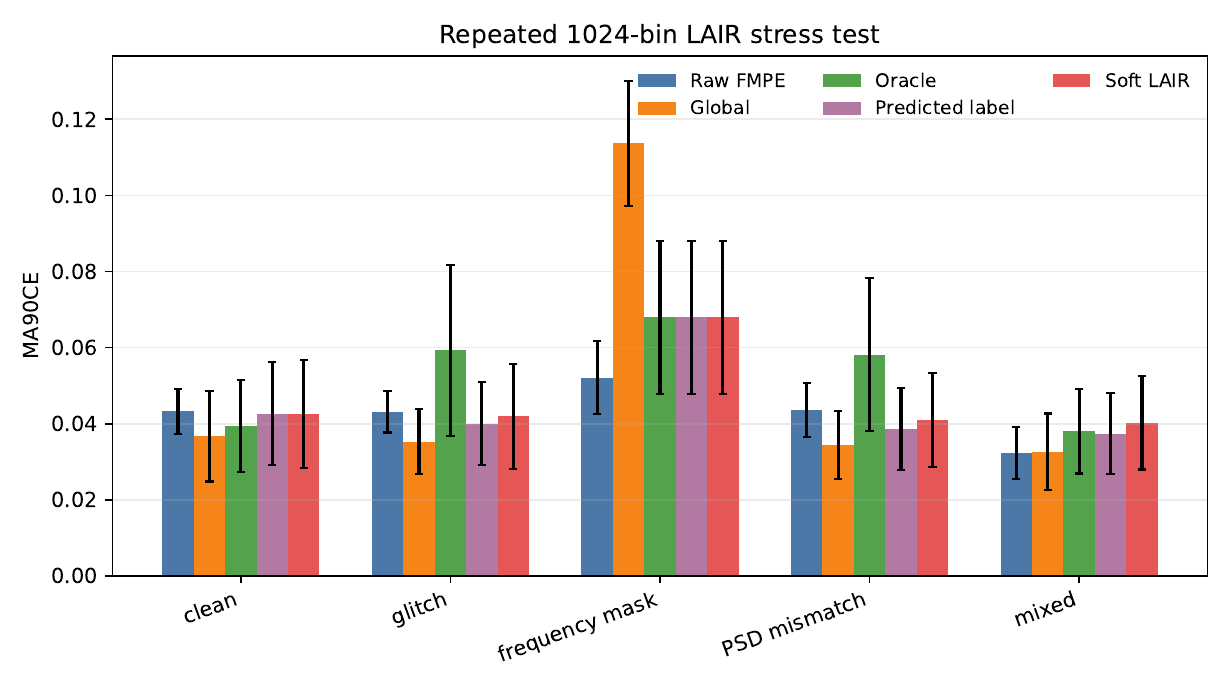}
		\caption{Repeated 1024-bin LAIR calibration study across 40 independent simulator/calibration splits. Error bars show split-to-split standard deviation.}
		\label{fig:lair-1024-multiseed}
	\end{figure}
	
	Table~\ref{tab:lair-1024-multiseed} and
	Fig.~\ref{fig:lair-1024-multiseed} report the repeated calibration study. The
	40 repeats used the validation-selected classifier from the
	classifier comparison, whose balanced accuracy was 0.5029; the study is
	therefore reported separately from the higher-performing classifier in
	Table~\ref{tab:classifier-1024}. The frequency-mask pattern persists: global
	rescaling has mean MA90CE 0.1137, while soft LAIR gives 0.0679. In mixed mode,
	raw FMPE, global rescaling, and soft LAIR have mean MA90CE 0.0323, 0.0327, and
	0.0403, respectively.
	
	\subsection{Artifact Classifier}
	
	\begin{table}[!htbp]
\centering
\caption{1024-bin artifact-classifier validation metrics on 2048 balanced validation events.}
\label{tab:classifier-1024}
\begin{tabular}{lr}
\toprule
Metric & Value \\
\midrule
Accuracy & 0.6982 \\
Balanced accuracy & 0.6982 \\
Negative log likelihood & 0.5028 \\
Expected calibration error, 10 bins & 0.0489 \\
Clean recall & 0.5918 \\
Glitch recall & 0.2070 \\
Frequency-mask recall & 1.0000 \\
PSD-mismatch recall & 0.9941 \\
Clean precision & 0.4951 \\
Glitch precision & 0.4753 \\
Frequency-mask precision & 1.0000 \\
PSD-mismatch precision & 0.7261 \\
\bottomrule
\end{tabular}
\end{table}

	Table~\ref{tab:classifier-1024} reports the 1024-bin artifact classifier. Its
	balanced accuracy is 0.6982, negative log likelihood is 0.5028, and expected
	calibration error is 0.0489. Frequency-mask recall is 1.0000 and PSD-mismatch
	recall is 0.9941. Clean recall is 0.5918, and glitch recall is 0.2070. Thus,
	for this classifier, masks and PSD mismatch are well identified, while glitch
	recognition is the main limitation for LAIR.
	
	The classifier comparisons test whether residual or focal-loss variants improve
	this tradeoff.
	
	\begingroup
\begin{center}
\refstepcounter{table}\label{tab:classifier-glitch-sweep}
{\scriptsize \textbf{TABLE~\thetable.} Glitch-targeted 1024-bin artifact-classifier comparisons. Bal. is balanced accuracy, NLL is negative log likelihood, and the recall columns correspond to clean, glitch, frequency-mask, and PSD-mismatch classes.\par}
\vspace{2pt}
\scriptsize
\setlength{\tabcolsep}{3pt}
\resizebox{\columnwidth}{!}{%
\begin{tabular}{lrrrrrr}
\toprule
Model & Bal. & NLL & $R_c$ & $R_g$ & $R_m$ & $R_p$ \\
\midrule
baseline & 0.6982 & 0.5028 & 0.5918 & 0.2070 & 1.0000 & 0.9941 \\
residual+focal & 0.5015 & 0.8247 & 0.1270 & 0.6289 & 1.0000 & 0.2500 \\
deep+focal & 0.5049 & 0.8241 & 0.0078 & 0.0605 & 1.0000 & 0.9512 \\
wide residual & 0.4980 & 0.8289 & 0.8164 & 0.0508 & 1.0000 & 0.1250 \\
validation seed & 0.7510 & 0.3815 & 0.7715 & 0.2539 & 1.0000 & 0.9785 \\
\bottomrule
\end{tabular}%
}
\end{center}
\endgroup

	Table~\ref{tab:classifier-glitch-sweep} gives the resulting tradeoff. In the
	initial three-configuration comparison, a residual width-96 focal-loss run raises
	glitch recall to 0.6289, but balanced accuracy falls to 0.5015, negative log
	likelihood increases to 0.8247, and clean and PSD-mismatch recall deteriorate.
	A follow-up residual/focal run with fresh validation seeds reaches balanced
	accuracy 0.7510 and negative log likelihood 0.3815. It improves clean,
	frequency-mask, and PSD-mismatch recognition, but glitch recall remains
	0.2539.
	
	\subsection{Prior and Gaussian Baselines}
	
	\begingroup
\begin{center}
\refstepcounter{table}\label{tab:gaussian-baseline}
{\scriptsize \textbf{TABLE~\thetable.} Diagonal Gaussian posterior baseline on the 256-bin support-aware low-resolution setting. Values are MA90CE.\par}
\vspace{2pt}
\scriptsize
\resizebox{\columnwidth}{!}{%
\begin{tabular}{lrrr}
\toprule
Mode & Prior only & Gaussian raw & Gaussian global \\
\midrule
clean & 0.0381 & 0.0313 & 0.0650 \\
glitch & 0.0706 & 0.0256 & 0.1181 \\
frequency mask & 0.0194 & 0.0288 & 0.0900 \\
PSD mismatch & 0.0175 & 0.0425 & 0.1094 \\
mixed & 0.0263 & 0.0219 & 0.0994 \\
\bottomrule
\end{tabular}%
}
\end{center}
\endgroup

	Table~\ref{tab:gaussian-baseline} illustrates the limits of a marginal coverage
	metric. The prior-only baseline has low MA90CE in several modes because broad
	central prior intervals often contain the truth. The diagonal Gaussian
	approximation also has low coverage error in some settings despite its limited
	posterior structure. Coverage error must therefore be considered together with
	posterior width and geometry.
	
	\subsection{Controlled-Likelihood Reference Checks}
	\label{sec:reference-posterior}
	
	A first check compared a quadratic Laplace approximation with FMPE samples for
	five clean 1024-bin events. The mean sliced Wasserstein distance was 0.3988
	and the median Hessian condition number was $2.119\times 10^5$, indicating a
	poorly conditioned local approximation rather than a reliable reference
	posterior. An exploratory importance-sampling run over three clean events
	produced median effective sample size 1.0 from 20000 prior proposals. These
	runs show that the controlled likelihood can be interrogated, but they do not
	provide validated reference posteriors.
	
	\begin{table*}[t]
\centering
\caption{Controlled-likelihood reference-posterior checks. $W_{\rm sliced}$ and $W_{\rm par}$ are mean sliced and parameter-wise Wasserstein distances in prior-unit coordinates. The dynesty and single-chain HMC rows did not satisfy the stated diagnostics. The final stretch-MCMC rows use a Laplace-preconditioned ensemble proposal and meet the split-$\hat R$ and effective-sample-size diagnostics for the target clean and frequency-mask event counts.}
\label{tab:dynesty-attempt}
\scriptsize
\begin{tabular}{llrrrr}
\toprule
Method & Mode & Ev. & $W_{\rm sliced}$ & $W_{\rm par}$ & diagnostic \\
\midrule
dynesty & clean & 5 & 0.4399 & 0.356 & median logzerr 0.668 \\
dynesty & frequency mask & 2 & 0.3851 & 0.347 & median logzerr 0.664 \\
HMC & clean & $3\times5$ & 0.3947--0.4492 & -- & median acceptance 0.0 \\
HMC & frequency mask & $2\times2$ & 0.3858--0.4297 & -- & median acceptance 0.0 \\
stretch MCMC & clean & 5 & 0.3912 & 0.3301 & $\max\hat R=1.0089$, ESS$_{\min}=3744$ \\
stretch MCMC & frequency mask & 2 & 0.4705 & 0.3596 & $\max\hat R=1.0091$, ESS$_{\min}=4182$ \\
\bottomrule
\end{tabular}
\end{table*}

	Table~\ref{tab:dynesty-attempt} reports further controlled-likelihood
	reference-posterior checks. The dynesty runs cover the target event counts,
	namely five clean and two frequency-mask 1024-bin observations, but they do
	not meet the evidence-error gate: the median evidence-error diagnostics are
	0.668 and 0.664 for clean and frequency-mask events, respectively. Repeated
	single-chain Hamiltonian Monte Carlo (HMC) runs also did not pass diagnostics,
	with median acceptance rate zero in all completed clean and frequency-mask
	runs.
	
	The most stable reference check uses a Laplace-preconditioned ensemble stretch
	Markov chain Monte Carlo (MCMC) sampler on the same controlled Gaussian
	likelihood. For five clean events it gives
	mean sliced Wasserstein distance 0.3912, mean parameter-wise Wasserstein
	distance 0.3301, maximum split-$\hat R=1.0089$, and minimum effective sample
	size 3744. For two frequency-mask events the corresponding values are 0.4705,
	0.3596, 1.0091, and 4182. Postprocessing of the MCMC samples gives mean
	90\% interval-overlap ratios 0.0097 and 0.0109 for clean and frequency-mask
	events. The small overlaps indicate substantial differences between the
	trained FMPE posterior and the controlled-likelihood reference in these
	examples. This check is a simulator-level reference diagnostic; it is not a
	Bilby/PyCBC or LVK-production posterior comparison.
	
	\section{Discussion}
	
	The 1024-bin experiments show clear artifact dependence in marginal interval
	calibration. A global scale fitted on mixed calibration data reduces the error
	in some modes but substantially worsens frequency-mask coverage. Soft LAIR
	mitigates that failure, lowering the 1024-bin frequency-mask MA90CE from the
	global value of 0.1195 to 0.0672, but it does not dominate the raw FMPE
	intervals across all artifact modes. The repeated evaluation gives the same
	qualitative result across simulator and calibration seeds.
	
	Classifier behavior is part of this outcome. Frequency masks are explicit in
	the mask channel and are identified with high recall. PSD mismatch is also
	recognized reliably by the main classifier. Glitches are less well separated
	from clean observations in the evaluated feature representation. When the
	classifier assigns little probability to the glitch class, the soft LAIR scale
	is pulled toward other artifact scales, limiting glitch-conditioned rescaling.
	
	The auxiliary diagnostics constrain the interpretation of coverage. The
	waveform-resolution test shows that coarse frequency grids can change the
	inference problem. The prior-only and Gaussian baselines show that low coverage
	error can arise from broad or structurally simple intervals. Posterior-geometry
	plots and controlled-likelihood reference checks further show that marginal
	coverage does not determine joint posterior structure or agreement with a
	likelihood-based target.
	
	The support-aware coordinate transform and circular phase representation remove
	two avoidable implementation artifacts: decoded samples leaving the prior
	domain and discontinuities at the phase boundary. In this controlled
	benchmark, the remaining limitations are artifact-dependent interval behavior,
	weak glitch recognition, and differences between FMPE samples and the
	controlled-likelihood reference. Extension to production gravitational-wave
	inference would require stronger transient-artifact classification,
	multivariate posterior diagnostics, full-budget training seeds, and comparison
	with Bilby/PyCBC or related likelihood-based analyses in an
	International Gravitational-Wave Network (IGWN)-compatible environment.
	
	\section{Conclusions}
	
	We evaluated flow-matching neural posterior estimation in a controlled
	gravitational-wave inverse problem with explicit artifact labels. The analysis
	focused on marginal interval calibration for clean observations, synthetic
	glitches, frequency masks, and PSD mismatch.
	
	The central empirical observation is mode dependence. In the 1024-bin
	evaluation, global marginal rescaling improves some modes but raises the
	frequency-mask MA90CE to 0.1195. Soft LAIR reduces that value to 0.0672, while
	remaining worse than raw FMPE in PSD-mismatch and mixed mode. The 40-split
	calibration study supports the same interpretation: frequency masks are a
	stable failure mode for global rescaling, and artifact-conditioned rescaling is
	useful primarily as a mode-resolved interval diagnostic.
	
	The classifier results identify a practical bottleneck. Frequency masks and
	PSD mismatch are recognized reliably by the main 1024-bin classifier, whereas
	glitch recall is low. The best residual/focal seed improves aggregate
	validation metrics, reaching balanced accuracy 0.7510, but its glitch recall
	is still 0.2539. More transient-sensitive features or architectures are needed
	before glitch-conditioned interval scaling can be relied on.
	
	The supporting checks are necessary for interpreting these numbers.
	Support-aware coordinates keep bounded parameters inside the prior domain, and
	the circular phase embedding removes the artificial boundary at $0$ and
	$2\pi$. Waveform-resolution tests motivate the use of 1024- and 2048-bin
	grids. The prior-only, Gaussian, conditional-flow, Laplace, dynesty, and
	multi-chain MCMC checks show that marginal coverage alone can be misleading
	when intervals are broad, joint structure is wrong, or the neural posterior
	differs from a likelihood-based reference.
	
	Within this controlled setting, LAIR provides a reproducible way to expose
	artifact-conditioned interval failures in neural posterior estimation. Moving
	to production gravitational-wave inference would require stronger glitch
	classification, multivariate posterior validation, and comparison with
	production-style likelihood-based analyses on more realistic detector data.

	\FloatBarrier
	\bibliographystyle{apsrev4-2}
	\bibliography{refs}

@article{Green2020Flows,
  title = {Gravitational-wave parameter estimation with autoregressive neural network flows},
  author = {Green, Stephen R. and Simpson, Christine and Gair, Jonathan},
  journal = {Physical Review D},
  volume = {102},
  number = {10},
  pages = {104057},
  year = {2020},
  doi = {10.1103/PhysRevD.102.104057},
  eprint = {2002.07656},
  archivePrefix = {arXiv}
}

@article{Dax2021Dingo,
  title = {Real-time gravitational-wave science with neural posterior estimation},
  author = {Dax, Maximilian and Green, Stephen R. and Gair, Jonathan and Macke, Jakob H. and Buonanno, Alessandra and Sch{\"o}lkopf, Bernhard},
  journal = {Physical Review Letters},
  volume = {127},
  number = {24},
  pages = {241103},
  year = {2021},
  doi = {10.1103/PhysRevLett.127.241103},
  eprint = {2106.12594},
  archivePrefix = {arXiv}
}

@inproceedings{Dax2023FMPE,
  title = {Flow Matching for Scalable Simulation-Based Inference},
  author = {Dax, Maximilian and Wildberger, Jonas and Buchholz, Simon and Green, Stephen R. and Macke, Jakob H. and Sch{\"o}lkopf, Bernhard},
  booktitle = {Advances in Neural Information Processing Systems},
  year = {2023},
  eprint = {2305.17161},
  archivePrefix = {arXiv}
}

@article{Lipman2022FM,
  title = {Flow Matching for Generative Modeling},
  author = {Lipman, Yaron and Chen, Ricky T. Q. and Ben-Hamu, Heli and Nickel, Maximilian and Le, Matt},
  journal = {arXiv preprint arXiv:2210.02747},
  year = {2022},
  eprint = {2210.02747},
  archivePrefix = {arXiv}
}

@article{Dax2023DingoIS,
  title = {Neural Importance Sampling for Rapid and Reliable Gravitational-Wave Inference},
  author = {Dax, Maximilian and Green, Stephen R. and Gair, Jonathan and P{\"u}rrer, Michael and Wildberger, Jonas and Macke, Jakob H. and Buonanno, Alessandra and Sch{\"o}lkopf, Bernhard},
  journal = {Physical Review Letters},
  volume = {130},
  number = {17},
  pages = {171403},
  year = {2023},
  doi = {10.1103/PhysRevLett.130.171403},
  eprint = {2210.05686},
  archivePrefix = {arXiv}
}

@article{Ashton2019Bilby,
  title = {Bilby: A user-friendly Bayesian inference library for gravitational-wave astronomy},
  author = {Ashton, Gregory and Huebner, Moritz and Lasky, Paul D. and others},
  journal = {The Astrophysical Journal Supplement Series},
  volume = {241},
  number = {2},
  pages = {27},
  year = {2019},
  doi = {10.3847/1538-4365/ab06fc},
  eprint = {1811.02042},
  archivePrefix = {arXiv}
}

@article{Biwer2019PyCBC,
  title = {PyCBC Inference: A Python-based parameter estimation toolkit for compact binary coalescence signals},
  author = {Biwer, C. M. and Capano, Collin D. and De, Soumi and Cabero, Miriam and Brown, Duncan A. and Nitz, Alexander H. and Raymond, V.},
  journal = {Publications of the Astronomical Society of the Pacific},
  volume = {131},
  number = {996},
  pages = {024503},
  year = {2019},
  doi = {10.1088/1538-3873/aaef0b},
  eprint = {1807.10312},
  archivePrefix = {arXiv}
}

@article{Talts2018SBC,
  title = {Validating Bayesian Inference Algorithms with Simulation-Based Calibration},
  author = {Talts, Sean and Betancourt, Michael and Simpson, Daniel and Vehtari, Aki and Gelman, Andrew},
  journal = {arXiv preprint arXiv:1804.06788},
  year = {2018},
  eprint = {1804.06788},
  archivePrefix = {arXiv}
}

@article{Lemos2023TARP,
  title = {Sampling-Based Accuracy Testing of Posterior Estimators for General Inference},
  author = {Lemos, Pablo and Coogan, Adam and Hezaveh, Yashar and Perreault-Levasseur, Laurence},
  journal = {arXiv preprint arXiv:2302.03026},
  year = {2023},
  eprint = {2302.03026},
  archivePrefix = {arXiv}
}

@article{Raymond2024RealNoise,
  title = {Simulation-based Inference for Gravitational-waves from Intermediate-Mass Binary Black Holes in Real Noise},
  author = {Raymond, Vivien and Al-Shammari, Sama and G{\"o}ttel, Alexandre},
  journal = {arXiv preprint arXiv:2406.03935},
  year = {2024},
  eprint = {2406.03935},
  archivePrefix = {arXiv}
}

@article{Mao2025Gaps,
  title = {Robust and scalable simulation-based inference for gravitational wave signals with gaps},
  author = {Mao, Ruiting and Lee, Jeong Eun and Edwards, Matthew C.},
  journal = {arXiv preprint arXiv:2512.18290},
  year = {2025},
  eprint = {2512.18290},
  archivePrefix = {arXiv}
}

@article{Wette2020SWIGLAL,
  title = {SWIGLAL: Python and Octave interfaces to the LALSuite gravitational-wave data analysis libraries},
  author = {Wette, Karl},
  journal = {SoftwareX},
  volume = {12},
  pages = {100634},
  year = {2020},
  doi = {10.1016/j.softx.2020.100634},
  eprint = {2012.09552},
  archivePrefix = {arXiv}
}

@article{LVK2023O3OpenData,
  title = {Open data from the third observing run of LIGO, Virgo, KAGRA and GEO},
  author = {{The LIGO Scientific Collaboration} and {The Virgo Collaboration} and {The KAGRA Collaboration}},
  journal = {The Astrophysical Journal Supplement Series},
  volume = {267},
  number = {2},
  pages = {29},
  year = {2023},
  doi = {10.3847/1538-4365/acdc9f},
  eprint = {2302.03676},
  archivePrefix = {arXiv}
}

@article{LVK2026O4bOpenData,
  title = {Open Data from LIGO, Virgo, and KAGRA through the Second Part of the Fourth Observing Run},
  author = {{The LIGO Scientific Collaboration} and {The Virgo Collaboration} and {The KAGRA Collaboration}},
  journal = {arXiv preprint arXiv:2605.27090},
  year = {2026},
  eprint = {2605.27090},
  archivePrefix = {arXiv}
}
	
\end{document}